\documentstyle[11pt,paspconf]{article}

\begin{document}

\title{Dormant black holes tell a story about 
the  evolution of  active  galactic nuclei}

\author{Paolo Salucci and  Ewa Szuszkiewicz,}
\affil{International School for Advanced Studies, SISSA, Via Beirut 2-4,
I-34013 Trieste, Italy}

\author{Pierluigi Monaco}
\affil{Institute of Astronomy, Madingley Road, CB3 0HA Cambridge, UK}

\author{Luigi Danese}
\affil{International School for Advanced Studies, SISSA, Via Beirut 2-4,
I-34013 Trieste, Italy}

\begin{abstract}
In our recent paper (Salucci et al. 1998) we have investigated 
the mass distribution function  of massive dark
objects  in galaxies, exploiting the available optical
and radio observations.
Under the assumption that
massive black holes  power active galactic nuclei, we have compared
the mass functions of massive dark objects and  
black holes  responsible for
the observed activity. We have found that a scenario with  a single short
burst per active galactic nucleus  is in a good  agreement with the
available data. 
Here we  summarize and discuss the main points of our study. 
\end{abstract}


\section{Introduction}

How and why active galactic nuclei (AGN)
 form and evolve is still, in many respects, mysterious.
From direct counts of AGN and from the intensity of the backgrounds
at high energies we can infer that the  activity in nuclei of galaxies
was much higher in the past than in the present  universe. If we accept the
paradigm that nuclear activity in galaxies is sustained
by accretion onto a massive black holes (BHs) (see Rees 1996 for a review), 
 then
the problem of the location and discovery of the remnants of such
past activity immediately arises. Recently, the number of detections of
massive dark objects (MDOs) in nuclei of inactive galaxies has increased 
 rapidly
(Magorrian et al.\ 1998;
van der Marel 1997; Ford et al. 1997;
for a review, see Kormendy and Richstone
1995), suggesting that we
are finding out the fossil of the past nuclear activity.
In the following we will assume that MDOs are BHs dormant after
a shining past.

\section{Mass function of massive dark objects }

The mass  of  MDOs in galaxy
centers can be evaluated by using  very high resolution
spectroscopy and photometry of nearby host galaxies.
Magorrian et al.\ (1998) have successfully exploited the very high
resolution of HST photometry together with high resolution
spectroscopy from ground of 36 elliptical (E)  and S0 galaxies to estimate
their MDO masses. In spite of the large scatter on the data,
 they  confirm the existence of a  relationship between
the mass of the hot galactic component $M_{sph}$
and the MDO/BH mass, $M_{BH}$, already suggested by Kormendy (1993),
although they also claim the presence of a significant scatter.
Following a different  and more indirect approach,
van der Marel (1998)
has analyzed the HST photometry of a sample of 46 early type galaxies.

As a first  step in the construction of MDOs mass function (MF) we
(Salucci et al. 1998) have estimated the MF of the hot
component of the normal galaxies,
by exploiting the luminosity functions (LF), the average fraction
of the total luminosity in
the spheroidal component of galaxies of different
morphologies and adopting mass-luminosity ratio:
$M/L_V\simeq 5.5\ h(L_V/L_{\star})^{0.25} $. 
 The overall shape of the MF
is relatively flat at $M < 10^{11} M_{\odot}$
and exhibits an exponential decline for $M >10^{11} M_{\odot}$.
At $M > 5\times 10^{10} M_{\odot}$
the mass function is dominated by spheroids in E and S0 galaxies.
The Sa/Sab galaxies exhibit an exponential decline of their
MF for $M > 10^{10} M_{\odot}$ and Sbc/Scd galaxies for $M> 10^8
M_{\odot}$.

Next, we have   used the distribution
of the $M_{BH}/M_{sph}$ ratio to infer the mass function of MDOs.
The average value
of $M_{MDO}/M_{sph}$ is still under debate and it ranges from
$\sim 10^{-2}$ (Magorrian et al.\ 1998)
 to $\sim
2\times 10^{-3}$ (Ho 1998). 
Due to observational
difficulties, the results so far obtained are expected to give only
hints on the true statistics.
However, a large scatter of the ratio $x=M_{MDO}/M_{sph}$,
at constant $M_{sph}$, has been found 
regardless of different assumptions and data analyses.
Our MF is shown in Figure 1.
The total mass density predicted by this distribution is
$\rho_{BH}\simeq 8.2 \times 10^5\ M_{\odot}/Mpc^3
\ (H_o/70)$, where $H_o$ is a Hubble constant. Most of the mass density
is ascribed to BHs in E and
S0 galaxies, 
and only a small fraction
is in BHs resident in late spiral types.

A different approach to  evaluation of MDO mass function relies
on the hypothesis that radio emission from the nuclei of
radio quiet galaxies is related to the mass of their MDOs.
As a  matter of fact, a strong trend
between MDO mass and nuclear radio luminosities has been
found by Franceschini et al. (1998).
Moreover, similar  correlation  is expected in the case of tiny
mass inflows, when advection dominated accretion flows (ADAF) occur
(e.g. Mahadevan 1997) or in the case of self absorbed
radio emission.  Thus, the radio LF of the  nuclear
emission of low-power spheroidal galaxies can be used to probe
the MDO mass function. We have  used it as a consistency check for the
MF derived from the optical spectroscopy and photometry (see Figure 1).

\section{Mass function of relic black holes } 

Soltan (1982)  showed that the AGN counts can provide
a meaningful lower limit to the mass density deposited in the
galaxy centres.
In Salucci et al. (1998) we have showed that the
mass function of relic  BHs (AMF) can be obtained
with simple hypothesis by exploiting the
information on the evolution of AGN and QSO LFs.
The luminosity functions and cosmic evolutions
are available now
for optically (see e.g. Pei 1995) and soft X--ray selected objects
(Hasinger, 1998). We  have taken  into account the contribution
to the  BH mass function coming from type 2 or obscured AGN, which
are possibly responsible for a major portion of the intensity
of the 2-50 keV X--ray background (HXRB). This hypothesis is strongly
supported by  optical identifications of
serendipitous sources detected by BeppoSAX in the 5-14 keV band
(Fiore et al.\ 1998). The existence of type 2 AGN
is postulated by unified models of AGN (see Antonucci 1993
for a comprehensive review), which are strongly supported
by many observational evidences ( see e.g. Granato et al 1997).
The resulting MF is shown in Figure 1.

\section{Comparison}

The comparison of the MF of the relic BHs to the MDO
MF is  used to cast light on the characteristics
of the evolution of the nuclear activity in galaxies.

\begin{figure}
\vspace{8cm}
\includegraphics{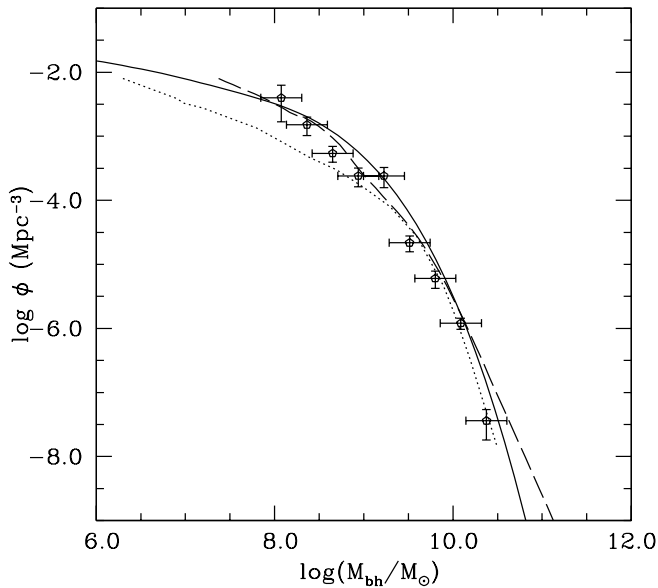}
\caption{
The mass function, $\phi_{OMF}d\log M_{BH}$,
derived using a Gaussian in the logarithm distribution of the ratio
$M_{MDO}/M_{sph}$ 
(solid line) compared to the mass function,
$\phi_{RMF}d\log M_{BH}$,
derived from the radio luminosity function of
of E/S0 cores (points with error bars).
and to the MF of
the relic BHs derived from the past activity of AGN
under assumption: $\lambda=(L/L_E)=10^{0.2(\log\ L-49)}$ (dashed line).
The dotted line is the MF of the relic BHs as
predicted by Cavaliere and Vittorini (1998).
}
\end{figure}

Nice agreement can
be found between the mass functions derived from investigations
on MDOs resident in local galaxies and the mass function
of the BHs inferred from the past activity of AGN
under very simple and sound hypotheses: {\it i)}
the nuclear activity is a single short event; {\it ii)} the spectra
of the AGN do not much depend on the redshift;
{\it iii)} the mass-radiation conversion efficiency
of accretion $\epsilon\simeq 0.1$; {\it iv)} the HXRB is produced by
absorbed AGN in the context of the (weak) unified scheme;
{\it v)}
$\lambda=L/L_E$ is an increasing function of the luminosity.
The last hypothesis is  strongly supported by observational
findings (see e.g. Padovani 1989). Recurrency or supply--limited accretion
at low luminosities may explain this result.


\begin{references}


\reference Antonucci R., 1993, ARA\&A, 31, 473






\reference Cavaliere A., Vittorini V., 1998, in The Young Universe;
Galaxy Formation and Evolution at Intermediate and High Redshift,
eds. S. D'Odorico, A. Fontana, E. Giallongo,
Astron. Soc. Pac. Conf. Ser. 146, 26









\reference Fiore F. et al. 1998, Nature, submitted 

\reference Ford H. C., Tsvetanov Z.I., Ferrarese L., Jaffe W.
1997,  in
Proceedings IAU Symposium 186, Kyoto, August 1997, D.B. Sanders,
J. Barnes, eds., Kluwer Academic Pub.

\reference Franceschini A., Vercellone S., Fabian A.C. 1998, 
MNRAS 297, 817

\reference Granato G.L., Danese L., Franceschini A. 1997, \apj, 486, 147


\reference Hasinger G., 1998, Astr. Nachtr. 319, 37


\reference Ho L.C., 1998,  in Observational
Evidence for Black Holes in the Universe, ed. S.K. Chakrabarti,
Kluwer Academic Pub.



\reference  Kormendy J., 1993, in The Nearest Active Galaxies, eds. J. Beckman,
L. Colina, \& H. Netzer (CSIC Press, Madrid), 197

\reference  Kormendy J., Richstone D., 1995, ARA\&A, 33, 581



\reference  Magorrian J., Tremaine S., Richstone D., Bender R.,
Bower G., Dressler A., Faber S.M., Gebhardt K., Green R.,
Grillmair C., Kormendy J., Lauer T.R., 1998, AJ, 115, 2285

\reference Mahadevan R., 1997, ApJ, 477, 585





\reference Padovani P., 1989, A \& A, 209, 27

\reference Pei Y. C., 1995, ApJ, 438, 623



\reference Rees M.J., 1996, in Black Holes and Relativity, ed. R. Wald,
Chandrasekhar Memorial Conference, Dec.




\reference Salucci P., Szuszkiewicz, E., Monaco, P., and
Danese, L.  1998, submitted to MNRAS 





\reference Soltan A., 1982, MNRAS, 200, 115


\reference van der Marel R.P., 1997,  in
Proceedings IAU Symposium 186, Kyoto, August 1997, D.B. Sanders,
J. Barnes, eds., Kluwer Academic Pub.

\reference van der Marel R.P., 1998, preprint astro-ph 9806365, submitted
to AJ







\end{references}
\end{document}